\documentclass[review,10pt]{JMtemplate}
\usepackage{bm}
\usepackage{times}
\usepackage{graphicx}
\usepackage{paralist,algorithmic,algorithm}
\usepackage{amsmath,mathrsfs,amssymb}
\usepackage{amsfonts}       
\usepackage{nicefrac}       
\usepackage{microtype}      
\usepackage{booktabs}       
\usepackage{longtable}
\usepackage{threeparttable}
\usepackage{multirow}
\usepackage{multicol}
\usepackage{soul}
\usepackage{subfigure}
\usepackage{wrapfig}
\usepackage[american]{babel}
\usepackage{algorithm}
\usepackage{algorithmic}
\usepackage{setspace}
\usepackage{stfloats}
\usepackage{lipsum}
\usepackage{multicol}
\usepackage{url}
\usepackage{makecell}   
\usepackage{hyperref}
\hypersetup{
    hidelinks,
    colorlinks=true,
    linkcolor=red,
    citecolor=blue,
    urlcolor=black
}
\usepackage{natbib}  
\usepackage{caption} 
\usepackage{appendix}

\newenvironment{proof}{{\noindent\it Proof.}\quad}{\hfill $\square$\par}

\begin{document}
\begin{frontmatter}
\title{Horizon-wise Learning Paradigm Promotes Gene Splicing Identification}

\author{Qi-Jie Li$^{1,2}$, \quad Qian Sun$^{1,3}$, \quad Shao-Qun Zhang$^{1,3,}$\footnotemark[1] \footnotetext[1]{Shao-Qun Zhang is the corresponding author. Email: zhangsq@lamda.nju.edu.cn} \\~\\
$^1$ \textit{National Key Laboratory for Novel Software Technology, Nanjing University, China} \\
$^2$ \textit{School of Artificial Intelligence, Nanjing University, Nanjing 210063, China} \\
$^3$ \textit{School of Intelligent Science and Technology, Nanjing University, Suzhou 215163, China}
}

\begin{abstract}
Identifying gene splicing is a core and significant task confronted in modern collaboration between artificial intelligence and bioinformatics. Past decades have witnessed great efforts on this concern, such as the bio-plausible splicing pattern AT-CG and the famous SpliceAI. In this paper, we propose a novel framework for the task of gene splicing identification, named Horizon-wise Gene Splicing Identification (H-GSI). The proposed H-GSI follows the horizon-wise identification paradigm and comprises four components: the pre-processing procedure transforming string data into tensors, the sliding window technique handling long sequences, the SeqLab model, and the predictor. In contrast to existing studies that process gene information with a truncated fixed-length sequence, H-GSI employs a horizon-wise identification paradigm in which all positions in a sequence are predicted with only one forward computation, improving accuracy and efficiency. The experiments conducted on the real-world Human dataset show that our proposed H-GSI outperforms SpliceAI and achieves the best accuracy of 97.20\%. The source code is available from \href{https://github.com/wnqn1597/H-GSI-}{this link}.
\end{abstract}

\end{frontmatter}

\section{Introduction}\label{introduction}
Recent years have witnessed a rapid development of artificial intelligence (AI) fueling scientific exploration and discovery due to the high accuracy and efficiency of AI techniques. Great efforts have been made to solve bioinformatics problems, such as AlphaFold \cite{alphafold}, Med-PALM \cite{palm}, and DNABERT \cite{dnabert}. Gene splicing is one of these difficult and significant tasks, which occurs between transcription and translation, retaining the exons and cutting out the introns \cite{rna-splicing}. Apposite splicing contributes to protein folding, whereas incorrect splicing may lead to the production of wrong proteins and ultimately affect the characterization of the organism, even serious and fatal diseases such as lung cancer and depression \cite{disease}.

\begin{figure}[t]
	\centering 
	\includegraphics[width=1\linewidth]{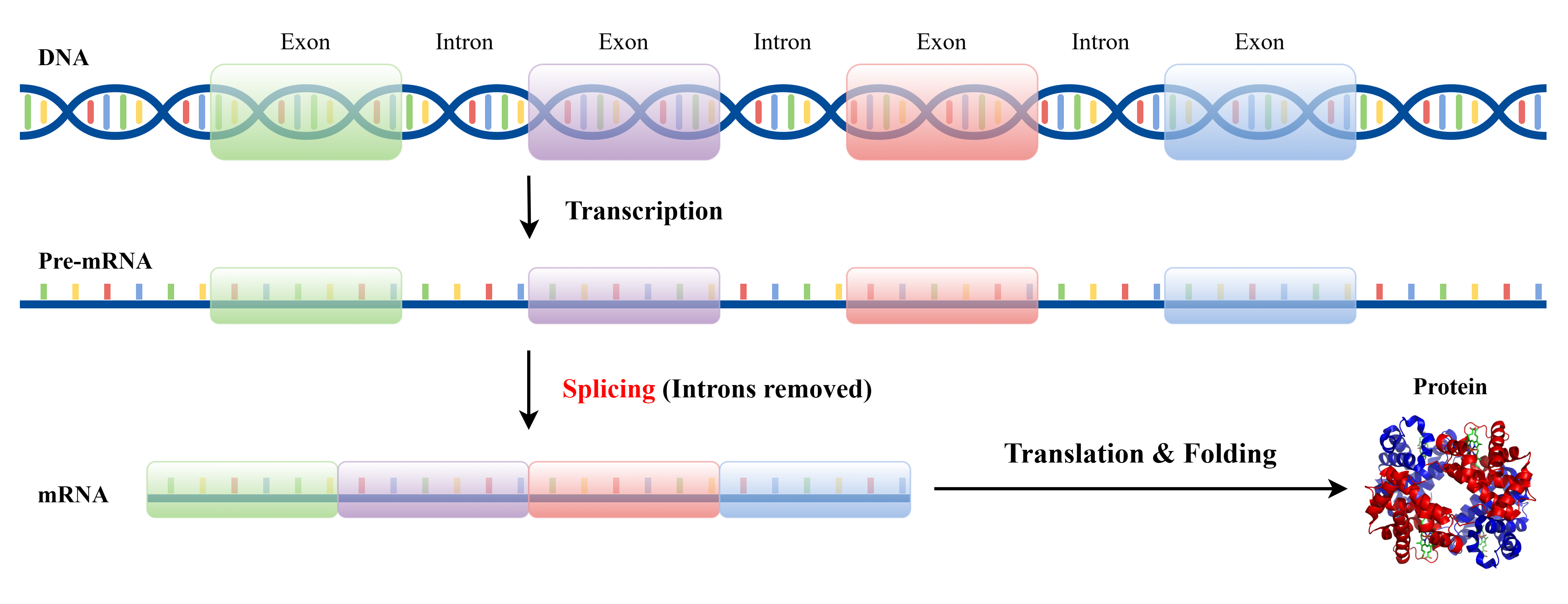}
	\caption{Graphical illustrations of gene expression procedure, which consists of these crucial steps: transcription, splicing, translation, and folding. Splicing occurs between transcription and translation, retaining the exons and cutting out the introns.} \label{fig:splicing}
\end{figure}

The gene sequence consists of four nucleotides, in which DNA contains the Adenine (A), Cytosine (C), Guanine (G), and Thymine (T) while RNA includes the first three and Uracil (U) instead of T. Eukaryotic genes contain many alternating exons and introns \cite{exon-intron}, which correspond to coding regions and non-coding regions, respectively. The gene expression procedure that uses the gene information to synthesize a functional product (mostly proteins) and ultimately affects a phenotype usually comprises four crucial steps: transcription, pre-mRNA splicing, translation, and folding, as illustrated in Figure \ref{fig:splicing}. The DNA is transcribed to produce the precursor messenger RNA (pre-mRNA) during transcription. Splicing occurs either during or immediately after transcription. It works by retaining the exon sequences to form mature messenger RNA while removing the introns \cite{splice-proc}, where splice sites are the boundaries between exons and introns. The mature RNA is usually the final gene product \cite{mrna}, each triplet nucleotide of which will be translated into an amino acid, determining the linear order of amino acids in a protein. Therefore, the precision of gene splicing identification determines the correctness of protein structure.

Traditional approaches for gene splicing identification are often established upon biological experiments testing and followed by statistical analysis of the observed splice sites \cite{stat1, stat2}. Two highly conserved dinucleotides GT-AG are observed on the splice sites and account for 98.71\% of all the splice sites \cite{stat1}. Regarding the GT-AG pattern as an important feature of gene splicing identification, there are still 0.56\% non-canonical splice sites, which make sense in biological events like immunoglobulin gene expression \cite{events}, and 0.73\% of remaining small groups \cite{stat1}. A huge volume of genome annotations shows that the majority of gene sequences that satisfy the GT-AG rule are not splice sites, thus statistical analysis is insufficient for accurate identification of splice sites. Besides, experimental methods for gene splicing identification are costly and time-consuming \cite{deepss}.

In past decades, many computational methods have been proposed, mostly based on classical machine learning models. These methods mostly follow a two-stage strategy: feature extraction and classification. Feature extraction is often achieved by combining features or transforming features to other domains via learning models, such as the Hidden Markov Models (HMMs) \cite{hmm:1,hmm:genesplicer:dt}, Distance Measure \cite{svm:3}, and FDTF encoding \cite{svm:1}. Then, various classification models like Support Vector Machines (SVMs) \cite{svm:splicemachine,svm:1,svm:2,svm:3} have been applied to give their predictions on encoding data and discriminate between true and false splice sites. Sonnenburg et al. employed SVM along with heterogeneous kernels and achieved 0.54 AUC-PR on the \textit{Homo sapiens} dataset \cite{svm:2}.

In recent years, some researchers have exploited deep-learning-based methods for handling this concern as the hotwave of deep learning. Existing methods usually include Recurrent Neural Networks (RNNs) \cite{rnn:2,rnn:splicevinci,rnn:1} and Convolutional Neural Networks (CNNs) \cite{deepsplicer, deepss, spliceai, splicefinder, splicecannon,deepsplice,splicerover}. CNNs use convolutional layers to extract local information, while RNNs use ``memory'' from prior elements within the sequence to grab global information. These models utilize hierarchical architectures to represent global high-level abstract features from the raw data and encapsulate highly complicated functions in the process. SpliceCannon \cite{splicecannon} obtained over 0.96 accuracy on the \textit{Homo sapiens} dataset. SpliceAI \cite{spliceai} achieved 0.95 top-k accuracy on the GENCODE dataset, which is a remarkable advance compared to conventional computational methods.

Existing methods mostly follow the \textit{point-wise identification paradigm}, which is fed up with a truncated fixed-length sequence and only predicts whether the midpoint of this sequence is a splice site. One who wants to use this model to predict splice sites for an unknown sequence of length $n$ must do forward computation for $n$ times, leading to an unbearably high cost. Besides, the sequence length processed by these methods is usually limited~\cite{zhang2020:hrp}, mostly no more than 400 nt. It is too short for an expressive model to extract long-range dependency from long sequences; thus, the accuracy is unsatisfactory. Furthermore, one-hot encoding is frequently utilized, which cannot capture local context information and imposes limitations on accuracy.

\begin{figure}[t]
	\centering
	\includegraphics[width=1.0\linewidth]{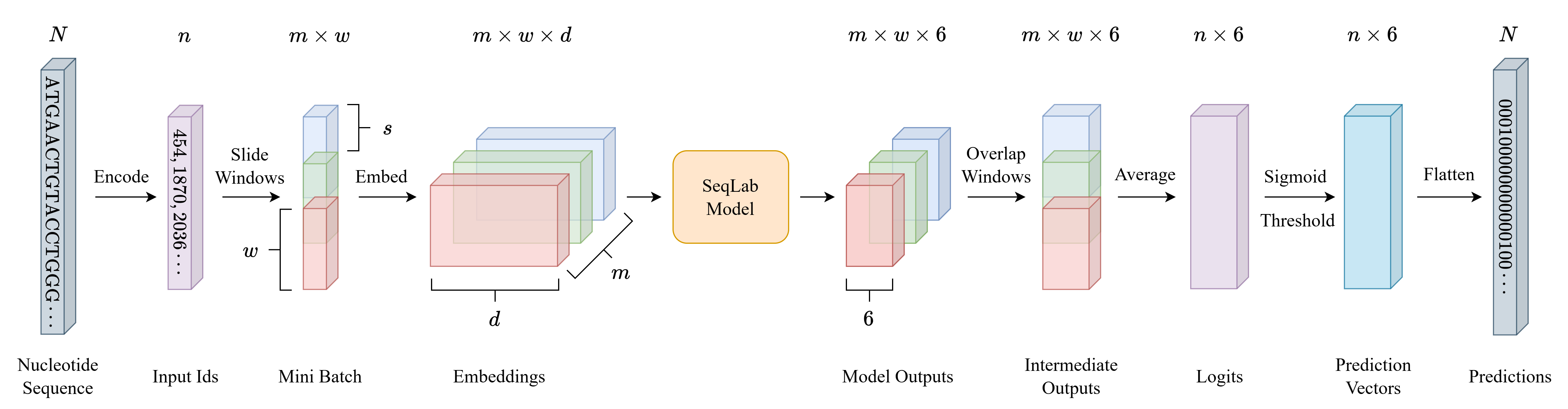}
	\caption{The workflow of our method H-GSI. Firstly, A nucleotide sequence of length $N$ is encoded into an integer vector of length $n$. Then the sliding window technique is applied and gives $m$ windows of size $w$ as a mini-batch. Embeddings of dimension $d$ are retrieved using integer vectors, as the input of the SeqLab model. The outputs are 6-dimensional real vectors indicating the logits. As a solution for overlapped outputs, the logits in overlapped segments are averaged. The prediction vectors are calculated via the sigmoid function and determined via fixed or dynamic thresholds. Finally, inference results are flattened from prediction vectors.} \label{fig:framework}
\end{figure}

In this paper, we propose a novel framework for the gene splicing identification task, named Horizon-wise Gene Splicing Identification (H-GSI). In contrast to existing studies that process gene information with a truncated fixed-length sequence, our proposed H-GSI follows the \textit{horizon-wise identification paradigm}, where all positions within the horizon of a sequence are predicted with only one forward computation. The proposed H-GSI comprises four components: the pre-processing procedure that utilizes 6-mer encoding to capture more local context information, the sliding window technique that handles long sequences without dropping global information, the SeqLab model structure, and the predictor, as illustrated in Figure~\ref{fig:framework}. The empirical investigations on the real-world Human dataset show that our proposed H-GSI outperforms competitors, achieving the best accuracy of 97.20\%, even under extremely class-imbalanced circumstances. 

The rest of this paper is organized as follows. Section \ref{related} reviews related studies. Section \ref{methodology} presents our methodology. Section \ref{experiments} conducts experiments on real-world datasets. Section \ref{conclusion} concludes this paper with discussions and prospects.

\section{Related Works}\label{related}

Many statistical analyses have been done to figure out the pattern used in gene splicing identification. Burset et al. \cite{stat1} extracted annotated examples of splice sites from the genomic databases, generated recognition weight matrices or consensus sequences, and analyzed possible splicing mechanisms based on conserved regions of non-canonical splice sites. Sheth et al. \cite{stat2} classified the splice sites into various categories and applied phylogenetic approaches to infer the evolutionary histories of the different splice-site subtypes.

There are several methods based on classical machine learning models, including SVMs~\cite{svm:splicemachine, svm:1, svm:2, svm:3}, HMMs~\cite{hmm:1, hmm:genesplicer:dt}, logistic regressions~\cite{lr}, decision trees~\cite{hmm:genesplicer:dt, zheng2023:CART}, random forests~\cite{rf}, restricted Boltzmann Machines~\cite{rbm}, and multilayer perceptrons~\cite{splicevec}. GeneSplicer~\cite{hmm:genesplicer:dt} applied a decision tree algorithm and enhanced it with Markov models to capture additional information around splice sites. SpliceMachine \cite{svm:splicemachine} employed a linear SVM to compute a linear classification boundary, with each candidate splice site represented as a feature vector containing its context information. Pashaei et al. employed an Adaboost classifier and achieved higher accuracy than several previous SVM-based methods \cite{ada}. Lee and Yoon proposed a novel deep belief network with a Restricted Boltzmann Machines training method for class-imbalanced splice site prediction \cite{rbm}. 

With the advance of computational techniques, deep learning models have achieved prominent success in recent times, among which RNNs and CNNs are mostly utilized in this task. SpliceFinder \cite{splicefinder} utilized a CNN solely for model training to deal with canonical and non-canonical splice sites. Zhang et al. \cite{splicecannon} firstly computed k-nucleotide frequencies and encoded the sequences via the Doc2vec technique. Then they used their proposed model SpliceCannon, which includes bidirectional long short-term memory networks and Convolutional Block Attention Module, to do subsequent prediction.

\section{Methodology}\label{methodology}

\begin{figure}[t]
    \centering
    \includegraphics[width=0.6\linewidth]{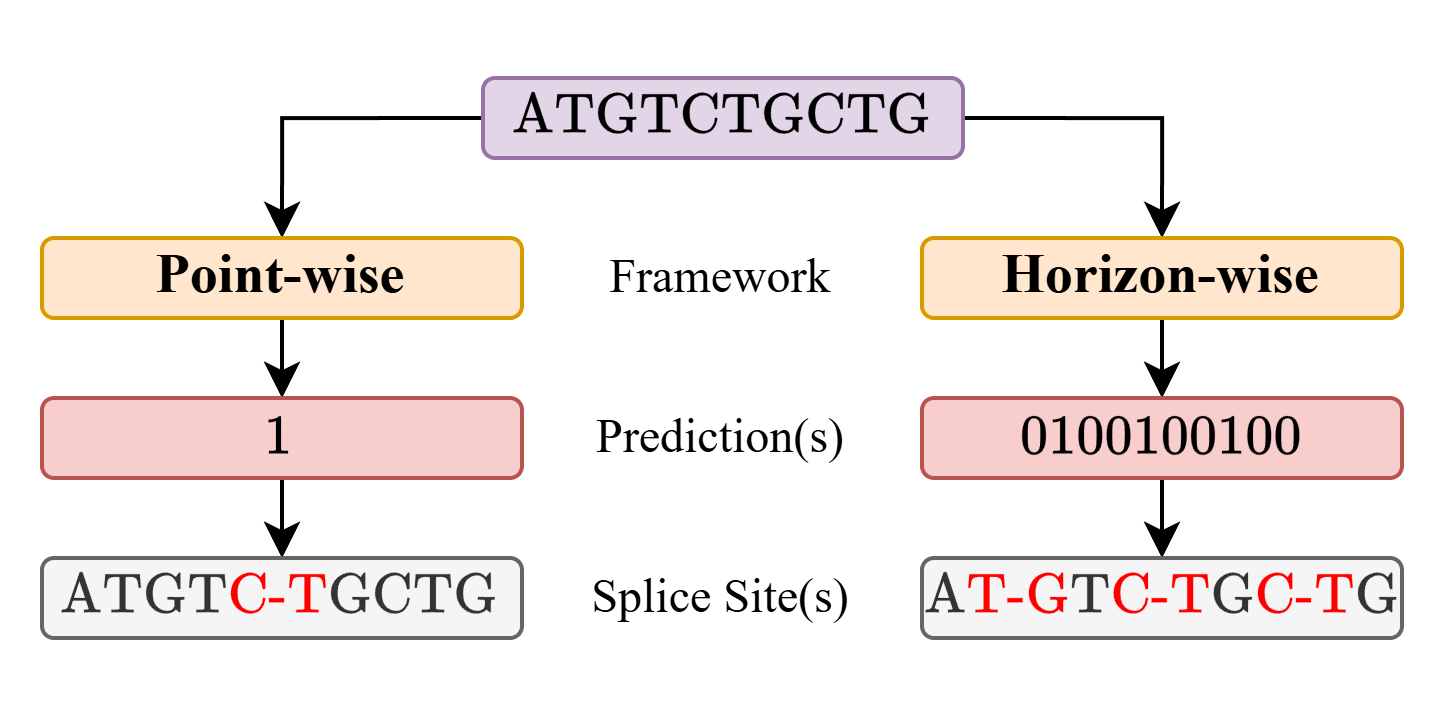}
    \caption{The difference between the point-wise identification paradigm and the horizon-wise identification paradigm.}
    \label{fig:paradigm}
\end{figure}

In the task of gene splicing identification, a framework $f$ takes a DNA sequence $S$ as input and produces an output $L$, that is, $f:\mathcal{S}\to\mathcal{L}$, where $\mathcal{S}$ is the set of DNA sequences, $\mathcal{L}$ is the set of labels, $S \in \mathcal{S}$, and $L \in \mathcal{L}$.

Existing identification methods often employ the point-wise identification paradigm, where the label indicates whether the midpoint of the sequence is a splice site. Let a finite set of symbols $\Gamma=\{\mathrm{A, C, G, T}\}$ denote these four nucleotides in DNA. A DNA sequence of length $2M$ is defined as a string $S=\left\langle s_{-M},\dots,s_{-1},s_1,\dots,s_M\right\rangle$, where $s_i\in\Gamma$. The corresponding label is an indicator variable $l\in\{0,1\}$, where $l=1$ declares a splice site between $s_{-1}$ and $s_1$. Taking an example of Figure \ref{fig:paradigm}, a sequence ``ATGTCTGCTG''  and a corresponding label ``1'' declare a splice site between C and T. However, the point-wise identification paradigm works costly since it only predicts whether the midpoint of a sequence is a splice site with one forward computation.

Our work employs the horizon-wise identification paradigm, where the label records whether each position in the horizon is a splice site. Let $N$ denote the horizon size. A DNA sequence of length $N$ is defined as a string $S=\left\langle s_1,s_2,\dots,s_N\right\rangle$, where $s_i\in\Gamma$. The corresponding label is also a string with the form of $L=\left\langle l_1,l_2,\dots,l_N\right\rangle$, where $l_i\in\{0,1\}$, where $l_i=1$ means that the right side of $s_i$ is a splice site. Taking the same example, the label ``0100100100'' declares three splice sites depicted in Figure \ref{fig:paradigm}. The horizon-wise identification paradigm gives predictions for all positions in a sequence together, thus achieving higher efficiency.

In the rest of this section, we formally propose our method H-GSI, which consists of four components: the pre-processing procedure, the sliding window technique, the SeqLab model, and the predictor. Figure~\ref{fig:framework} illustrates the workflow of H-GSI.

\subsection{Pre-processing Procedure}
The raw data of gene sequences are initially in string form and incompatible with computational models. Thus, it is necessary to transform the sequences into tensor form during the pre-processing procedure. Existing methods usually adopt the one-hot encoding technique, which transforms categorical labels into vectors of 0 and 1. Each dimension of a vector corresponds to a particular nucleotide type. Thus, a DNA sequence of length $n$ is transformed into a $n\times4$ matrix. For example, a sequence ``ACGTA'' is encoded as a $5\times4$ matrix
\begin{equation}  \label{eq:one-hot}
\begin{bmatrix}
\mathrm{A}\\
\mathrm{C}\\
\mathrm{G}\\
\mathrm{T}\\
\mathrm{A}
\end{bmatrix}
\to
\begin{bmatrix}
1&0&0&0\\
0&1&0&0\\
0&0&1&0\\
0&0&0&1\\
1&0&0&0
\end{bmatrix}
\end{equation}

However, a nucleotide sequence is rich in biomolecular messages, such as sequence modifications and hydrogen bonding. From Eq.~\eqref{eq:one-hot}, only one nucleotide per position leads to the encoded sequence being too long so that the model can hardly extract global information. Thus, one-hot encoding not only discards local context information but also reduces efficiency.

\begin{figure*}[t]
	\centering
        \subfigure[Encoding]{
            \label{fig:encode}
            \includegraphics[width=0.48\linewidth]{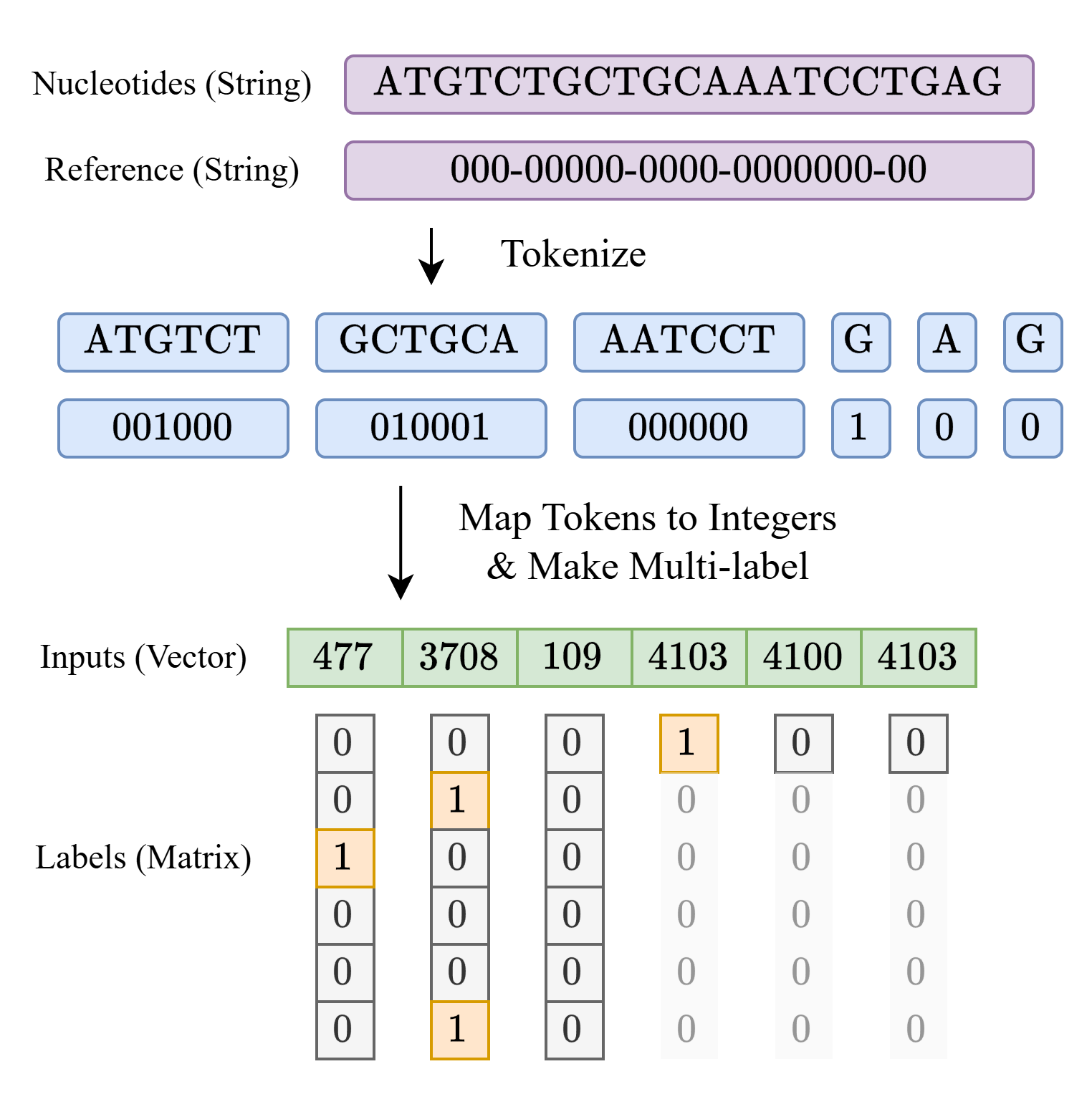}
        }
        \subfigure[Prediction]{
            \label{fig:predict}
            \includegraphics[width=0.48\linewidth]{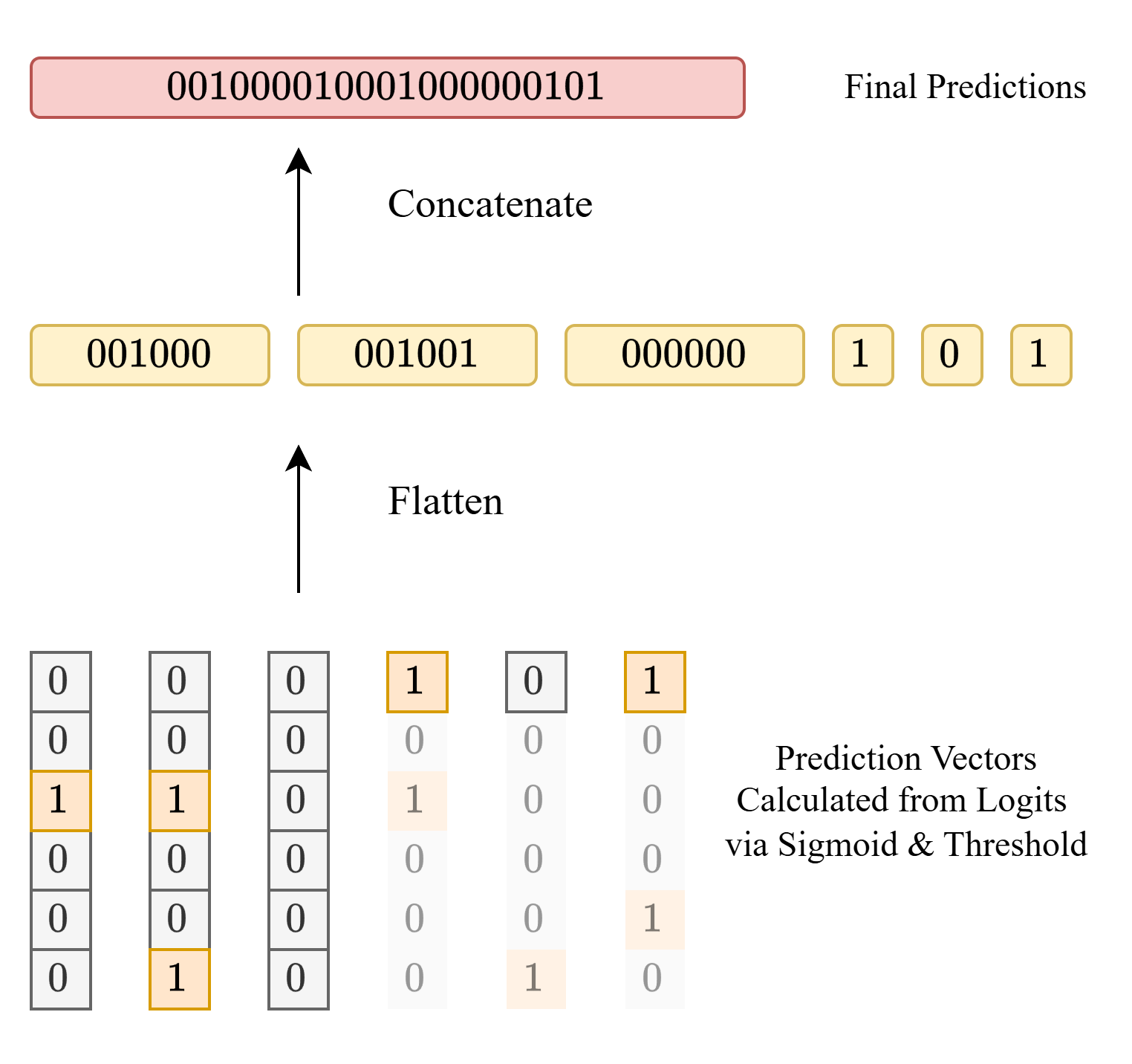}
        }
	\caption{Illustrations of (a) Encoding and (b) Prediction.} \label{fig:preprocess}
\end{figure*}

To tackle these challenges, we propose an alternative encoding method in the pre-processing procedure. A nucleotide sequence is transformed into an integer vector via a six-mer tokenizer. More specifically, a group of six contiguous nucleotides is mapped to an integer. If the number of rest nucleotides is less than six, then the tokenizer maps each nucleotide to a single token. Each token has a length of either six or one. Taking an example of Figure~\ref{fig:encode}, a nucleotide sequence in string format is split into an ordered list of tokens and mapped to an integer vector as input of the model.

The setting of the six-mer tokenizer is bio-inspired. During translation, three nucleotides constitute a codon, and the type of amino acid is determined \cite{codon}. Therefore, it is natural that one forces the token length to equal the multiple of three. The three-mer token may be slightly short and does not have sufficient capacity to capture local information, while the nine-mer tokenizer would result in a huge vocabulary size ($4^9=262144$). Therefore, we utilize the six-mer tokenizer to strike a balance between expressiveness and efficiency. Besides, the mapping keys and values are defined in the vocabulary file. We do not use the special token ``CLS'' since the sequence is a segment intercepted from the complete genome. The remaining space is padded by the special token ``PAD'' if the sequence length is less than the window size. The concept of window size is described in Subsection \ref{sliding-window}.

According to supervised learning, we also need to encode the labels of the reference sequence that correspond to the token input sequence. Here, we pre-process the original reference sequences described in Subsection~\ref{dataset}. Then we employ a multi-label scheme and collect six labels for each token, where each label represents whether the corresponding position is a splice site. Let $1\le i\le 6$, and $l_i\in\{0, 1\}$ denotes the $i$-th label, in which $l_i=1$ means that the right side of the $i$-th nucleotide is the splice site and $l_i=0$ means not. So we can convert a collection of six labels into an exact vector of $[l_1,\dots,l_6]$. Similar to the transformation in nucleotide sequences, if the number of rest labels is less than six, one marks the first element of the vector as $l_1$ and pads five extra ``0'' as placeholders~\cite{zhang2021:life}.

Taking the example of Figure \ref{fig:encode}, the nucleotide token ``ATGTCT" corresponds to the label token ``001000", where $l_3 = 1$ means that there is a splice site between G and T. For the case that one label token contains several splice sites, such as the nucleotide token ``GCTGCA" corresponds to the label token ``010001", $l_2=1$ and $l_6=1$ mean that there are two splice sites: one is between C and T, and the other is after A.

After the pre-processing procedure, both nucleotide and reference sequences in string format are successfully transformed into vectors compatible with machine-learning models.

\subsection{Sliding Window Technique}\label{sliding-window}
To deal with large-scale sequences, we also employ the sliding window technique, which contributes to slicing the token input sequences. Let $w$ and $s$ denote the fixed window size and the sliding stride, respectively. We perceive $w$ tokens inside a fixed window and slide the window to the right by $s$ tokens. A token input sequence of length $n$ consists of $\lceil\frac{n-w}{s}\rceil+1$ overlapped windows. For example, a sequence of $[477, 3708, 109, 4103, 4100, 4103] $ is transformed into a set of sequences 
\[
    \left\{ \ [477, 3708, 109], \ [109, 4103, 4100], \ [4100, 4103] \ \right\}
\]
when $w=3$ and $s=2$. In this paper, we set $w=1000$ and $s=500$ as the default.

\subsection{SeqLab Model}
The horizon-wise identification paradigm needs a sequence-handling model whose inputs and outputs are sequences of the same length. As illustrated in Figure~\ref{fig:framework}, the states before and after the model are all of sequential formats. The sequence labeling (SeqLab) model that takes sequential data as input and generates a categorical label for each sequence element naturally meets the requirement. There are a variety of SeqLab models, such as CNNs, RNNs, and Transformers. Equipped with SeqLab models, the proposed H-GSI can capture more structural information than independent classifiers. 

In this paper, we investigate the performance of various SeqLab models, including the Fully Convolutional Network (FCN), Temporal Convolutional Network (TCN) \cite{tcn}, Gated Recurrent Unit Model (GRU) \cite{gru}, Long Short-Term Memory Network (LSTM) \cite{lstm}, and Transformer \cite{attention}. The experiments are detailed in Section~\ref{experiments}.

\subsection{Predictor}\label{predictor}
For each token, the SeqLab model produces a 6-dimensional real vector as the output, each dimension of which indicates the logit of the corresponding nucleotide being a splice site. Notice that the windows may overlap with each other when stride $s$ is less than window size $w$. There would be more than one version of model outputs for overlapping segments, as shown by the intermediate outputs in Figure \ref{fig:framework}. To tackle this challenge, we use the average of these intermediate outputs as the integrated logits. With this operation, we ensure that predictions for overlapping segments are based on both the left and right sides and integrate contextual information comprehensively.

Hence, based on the logits we can calculate the output probabilities. Now we focus on one logit in one of the logit vectors for clarity. The corresponding probability $p$ is calculated from this logit via the sigmoid function. Here, we employ the binary cross entropy loss between the probability $p$ and the corresponding true label $l$ as follows
\begin{equation}
\ell(p, l) = l\log p + (1-l)\log(1-p) \ ,
\end{equation}
After the training procedure, H-GSI generates the corresponding prediction $y$ by setting a probability threshold $\alpha \in [0, 1]$ on the probability $p$, i.e., 
\begin{equation}
    y=\begin{cases}
        1 \ , & p > \alpha \ , \\
        0 \ , & p \leq \alpha \ .
    \end{cases}
\end{equation}
In this paper, we use both fixed and dynamic thresholds. For the fixed threshold, we set $\alpha=0.5$. For the dynamic threshold, we collect all the true splice sites in the training set and sort these sites in descending order of output probabilities. Hence, the minimal probability value in the sorting is exactly the dynamic threshold we require. If the model does not suffer from overfitting, the number of positions that are predicted as splice sites may approach the number of true sites in the test set, which meets our expectations.

The operations to calculate the probability $p$ and prediction $y$ mentioned above are element-wise. So now the predictions are still in the form of 6-dimensional vectors, and the final prediction is made by concatenating these vectors. Taking the example of Figure \ref{fig:predict}, tokens ``ATGTCT'', ``GCTGCA'', and ``AATCCT'' represent six nucleotides each, and their corresponding prediction vectors will be flattened and concatenated. Meanwhile, tokens ``G'', ``A'', and ``G'' represent single nucleotides each. Only the first element of their corresponding vectors will be reserved and appended to the tail of the final prediction. In contrast, others are ignored since their corresponding labels are placeholders and make no sense.

\section{Experiments}\label{experiments}
In this section, we conduct H-GSI on real-world Human splicing variants in competition with the state-of-the-art contenders.

\begin{figure}[t]
\centering
\includegraphics[width=0.6\linewidth]{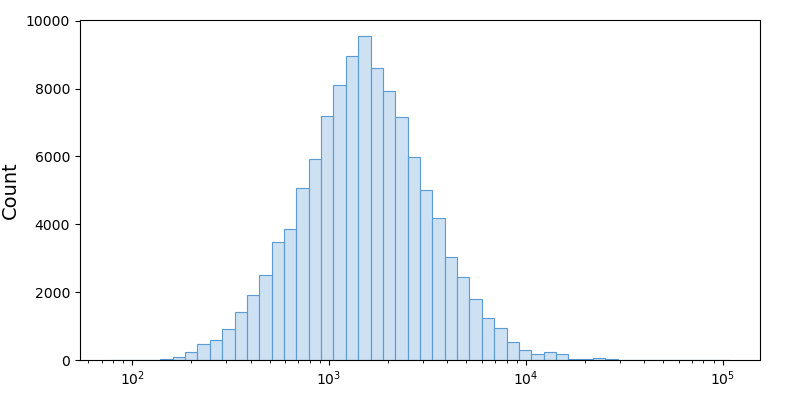} 
\caption{The histogram of sequence lengths in the Human dataset. The x-axis is log-transformed so that the distribution looks like Gaussian, but it is an intrinsically long-tailed distribution.}
\label{fig:hist-seq}
\end{figure}

\begin{figure}[t]
\centering
\includegraphics[width=0.6\linewidth]{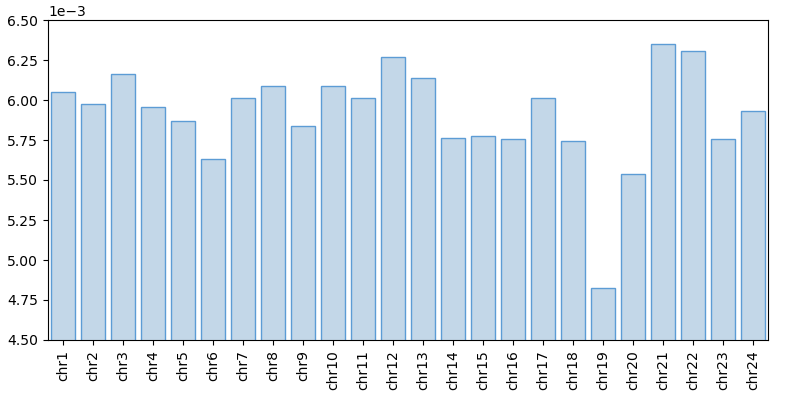} 
\caption{The ratio of splice sites to the whole sequence of each file in the Human dataset.}
\label{fig:ratio}
\end{figure}

\begin{figure}[t]
    \centering
    \subfigure[Base A]{
        \label{fig:A}
        \includegraphics[width=0.45\linewidth]{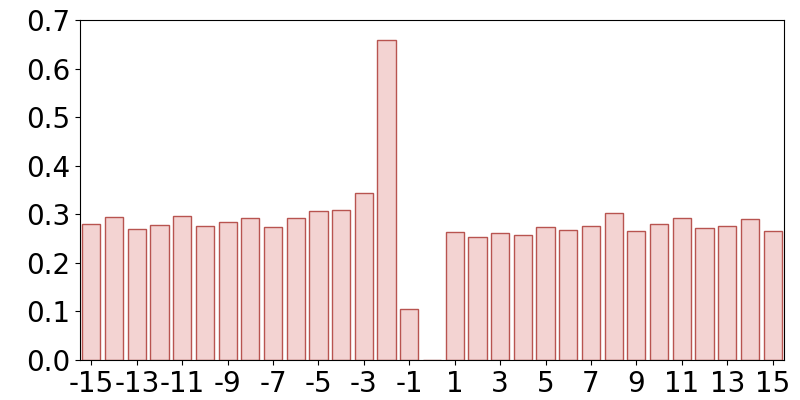}
    }
    \subfigure[Base T]{
        \label{fig:T}
        \includegraphics[width=0.45\linewidth]{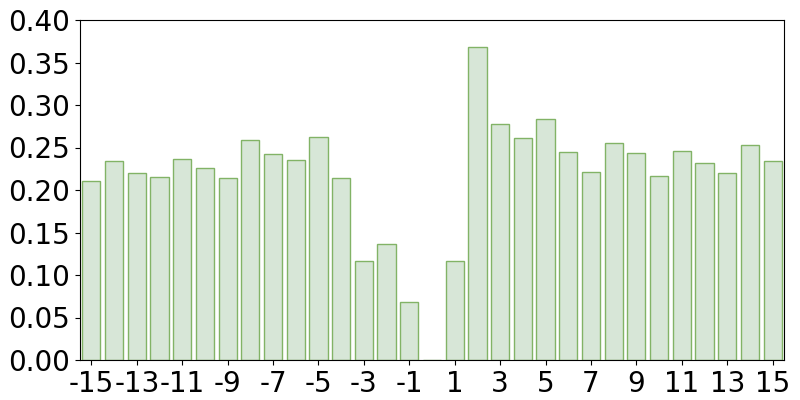}
    }
    
    \subfigure[Base C]{
        \label{fig:C}
        \includegraphics[width=0.45\linewidth]{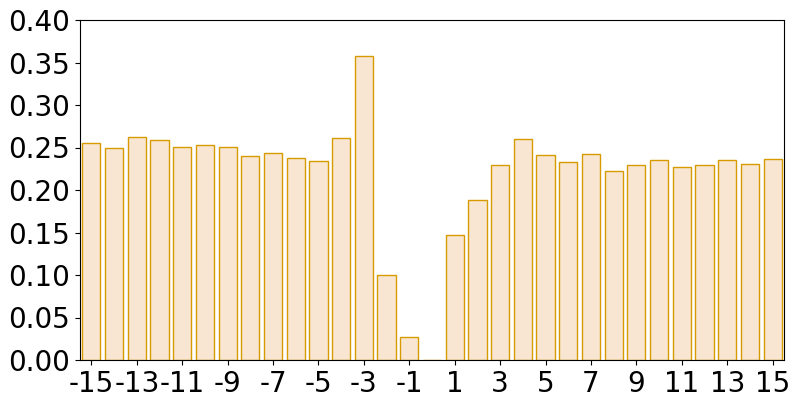}
    }
    \subfigure[Base G]{
        \label{fig:G}
        \includegraphics[width=0.45\linewidth]{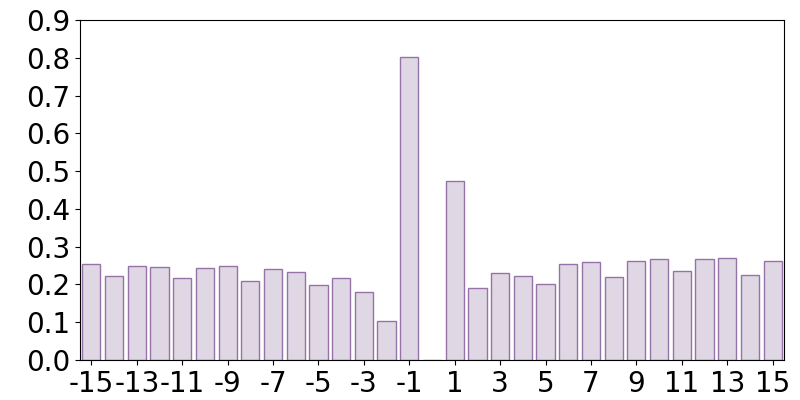}
    }
    \caption{The histograms of base A, T, C, and G at 15 positions around splice sites. The peaks around the origin note the GT-AG pattern, while the frequency of others is approximately equal to 0.25.}
    \label{fig:hist-15}
\end{figure}

\subsection{Human Dataset}\label{dataset}
Here, we focus on the Human dataset, which is a typical dataset of human splice sites and is publicly available at \href{https://grch37-archive.ensembl.org/info/website/index.html}{Ensembl Release 112}. The Human dataset comprises 110, 371 pieces of data, each consisting of a gene sequence and a reference sequence. One piece of data has varied lengths and contains about 2, 032 bases on average, of which the longest sequence consists of 107, 976 bases and the shortest sequence consists of only 78 bases. The gene sequence is composed of four bases (i.e., A, T, C, and G) and a splice site marker ``-'', where the splice site marker ``-'' represents a splice site. The histogram of sequence lengths is depicted in Figure \ref{fig:hist-seq}. The whole dataset contains four bases: A, T, C, and G, in which A, T, C, and G account for about 27.1268\%, 22.0656\%, 25.1311\%, and 25.6766\%, respectively. There are 1, 325, 155 splice sites in the dataset, which account for $0.5908\%$ of the entire sequence. 

The reference sequence is composed of ``0'' and the splice site marker ``-'', whose length is the same as its corresponding gene sequence, where the symbol ``0'' in the reference sequence matches a base in the corresponding gene sequence. Markers in gene sequence are redundant and discarded during pre-processing. Each marker in reference sequences and ``0'' on the left are transformed to ``1''. For example, the reference sequence ``000-000'' is transformed into ``001000''. The ratio of splice sites to the whole sequence of each file is shown in Figure \ref{fig:ratio}. We also calculated the frequency of base A, T, C, and G at 15 positions around splice sites depicted in Figure \ref{fig:hist-15}. On the one hand, these peaks near the origin are caused by the pattern GT-AG, while the ratio for each base is around one-quarter at other positions and does not show any specificity. On the other hand, most of the sequences that follow the GT-AG rule are not true splice sites.

Following the pre-processing procedure introduced in Section \ref{methodology}, we get the transformed dataset with six labels for one token. The positive samples for each label account for about 0.5572\%, 0.3818\%, 0.8290\%, 0.5661\%, 0.3843\%, and 0.8135\%, respectively. Thus, the transformed dataset is also extremely class-imbalanced.

\begin{figure}[t]
    \centering
    \includegraphics[width=0.65\linewidth]{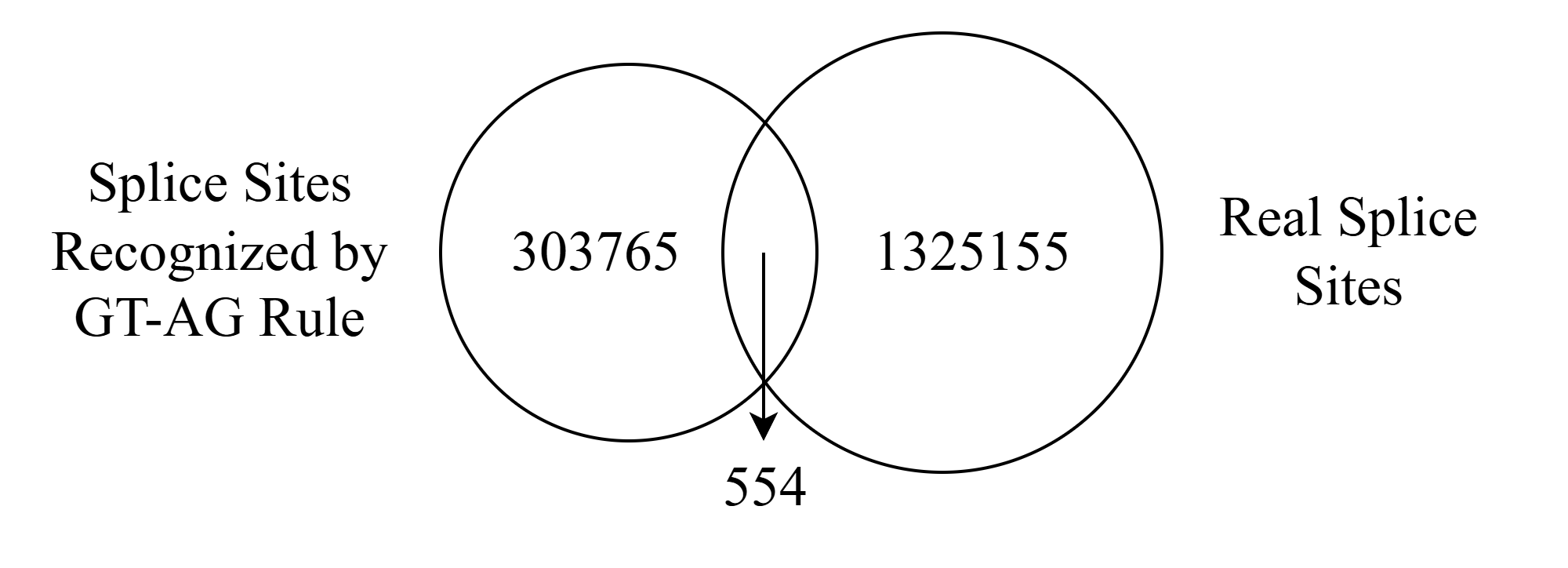}
    \caption{The relationship between splice sites recognized by the GT-AG rule and real splice sites in Human dataset, where the GT-AG rule recognizes 303, 765 splice sites, of which only 554 are true splice sites and the rest are discredited. There are 1, 325, 155 real splice sites, of which only 554 are recognized by the GT-AG rule and the rest are missed.}
    \label{fig:GT-AG}
\end{figure}

We also investigate the conventional GT-AG rule in the Human dataset. The empirical results are displayed in Figure \ref{fig:GT-AG}. It is observed that only a minority of splice sites are correctly identified via the GT-AG rule. Therefore, simple statistical analysis is not enough for gene splicing identification.

\subsection{Contenders and Configurations}

To elaborate on the effectiveness of H-GSI, we compare it with the state-of-the-art model SpliceAI-10k. SpliceAI-10k is a 32-layer fully convolutional network, which employs dilated convolution \cite{dilated} and residual block \cite{residual} for the increased receptive field of high-level neurons. SpliceAI-10k takes in long input (5k nt) to handle large chunks of nucleotide sequences and performs 3-class classification (no splice site, donor site, and acceptor site) per position. Here, we replace its classification head with 2-class (no splice site and splice site) because we only focus on the location of splice sites but not the category.

Section \ref{methodology} has introduced the requirements of H-GSI for the model architecture. In this experiment, we implement the proposed H-GSI equipped with various SeqLab models, including FCN, TCN, GRU, LSTM, and Transformer. 

The number of layers, kernel size, and dilation rate of the FCN are the same as those of SpliceAI-2k, which can handle 1k tokens (with 2k context). The channel size of FCN is chosen heuristically to ensure its competitive expression ability. The kernel size of TCN is 2 \cite{tcn}, and the dilation rate doubles with each layer. Therefore, we set the number of temporal convolution layers equal to 10 to ensure that the receptive field size of the last token is equal to $1024$, indicating that the prediction result of each token depends on its total precursor sequence. The channel size of TCN was chosen to be larger than that of FCN for comparable model parameter sizes. The architecture designs of GRU and LSTM are completely heuristic, only to guarantee the decrease in training loss is normal. The Transformer consists of alternating self-attention blocks and feed-forward layers. Here, we simply employ a fine-tuned ESM model \cite{esm}, which is a 500M parameter transformer pre-trained on the human reference genome \cite{model}. The detailed configurations of these learning models are listed in rows 2–6 of Table \ref{tab:para}.

\begin{table}[t]
    \caption{Configurations and hyper-parameters of H-GSI equipped with various models.}
    \label{tab:para}
    \begin{center}
    \begin{tabular}{l|rrrrr}
        \toprule[2pt]
         & FCN & TCN & GRU & LSTM & Transformer \\ 
        \midrule
        embedding dim & 32 & 32 & 512 & 512 & 1280 \\ 
        hidden dim & 128 & 384 & 1024 & 1024 & 1280 \\ 
        num layers & 24 & 11 & 4 & 4 & 24 \\ 
        dropout & 0 & 0 & 0 & 0 & 0.05 \\ 
        parameters size & 5.9M & 11.4M & 19.4M & 25.2M & 478.8M \\ 
        \midrule
        learning rate & $10^{-4}$ & $10^{-2}$ & $10^{-3}$ & $10^{-3}$ & $10^{-5}$ \\ 
        batch size & 32 & 32 & 32 & 32 & 32 \\ 
        epochs & 20 & 20 & 20 & 20 & 20 \\ 
        warmup steps & 0 & 0 & 0 & 0 & 1024 \\ 
        \bottomrule[2pt]
    \end{tabular}
    \end{center}
\end{table}

\subsection{Experimental Setup}
We randomly chose 10\% sequences of the total dataset as the test set and the remaining as the training set. It should be emphasized that we do not tune the hyper-parameters on a validation set; we just set them as the default. The hyper-parameters for training are shown in rows 7–10 of Table \ref{tab:para}. All training and evaluation processes are run with random seed 42.

All experiments were conducted on 8 GeForce RTX 3090 GPUs and 2 A100-80GB-PCIE GPUs. Both two servers are equipped with 1007GB memory and Ubuntu 20.04. The proposed method and baselines are implemented in PyTorch. The source code is available from \href{https://github.com/wnqn1597/H-GSI-1}{this link}.

\subsection{Evaluation Metrics}
\textit{Random Accuracy}. 
Normal accuracy metric which calculates the percentage of correctly classified elements is largely ineffective due to the property of unbalanced labels, i.e., most of the positions are not splice sites. If there is a model that always predicts a position as a non-splice site, it still achieves high normal accuracy, such as $99.4\%$, since the ratio of true splice sites is around $0.6\%$ shown in Figure \ref{fig:ratio}. Therefore, it is insufficient to use normal accuracy to measure the model's ability. 

To enhance the correlation between the accuracy and the ratio of predicted true splice sites, we should emphasize them more than non-splice sites. We can randomly reserve a small part of non-splice sites with a proportion $\gamma\in [0, 1]$ and discard other non-splice sites to improve the importance of true splice sites. Then we calculate the normal accuracy of processed labels and predictions. We call it random accuracy, which is more appropriate in such settings. The formal definition of random accuracy is described as follows. Let $\boldsymbol{y}=[y_1, \dots, y_N]$ denote the predictions and $\boldsymbol{l}=[l_1, \dots, l_N]$ the labels. Let $\mathbb I_i$ be the indicator of whether the prediction matches the label at position $i$, i.e., 
\begin{equation}
\mathbb I_i=\begin{cases}
1 \ , & y_i=l_i \ , \\
0 \ , & y_i\ne l_i \ .
\end{cases}
\end{equation}
Suppose that $P$ is the subscript set of positive labels (splice sites), and $N'$ is the subscript set of randomly retained negative labels (non-splice sites). The random accuracy is calculated as
\[
Acc = \frac{\sum_{i\in P}\mathbb I_i + \sum_{i\in N'}\mathbb I_i}{|P|+|N'|} \ .
\]
To reduce bias caused by stochasticity, we take the average of $r$ calculations as the final result. It is an unbiased estimation of the expectation of random accuracy. We claim that the expectation of random accuracy is equivalent to weighted accuracy\label{claim1}, as detailed in Appendix \ref{appendix}. This claim shows that the random accuracy metric strikes a balance between true and false splice sites by reassigning their weights. In this paper, we set $p=0.01$ and $r=10$ for evaluation, according to the label ratio reported in Subsection \ref{dataset}.

\textit{Precision \& Recall}.
The precision and recall metrics are separately calculated by
\[
Precision = \frac{TP}{TP+FP}
\quad\text{and}\quad
Recall = \frac{TP}{TP+FN} \ ,
\]
where $TP$, $TN$, $FP$, and $FN$ denote the number of correctly predicted splice sites, correctly predicted non-splice sites, mispredicted splice sites, and mispredicted non-splice sites, respectively.

\textit{Top-k Accuracy}.  Similar to SpliceAI \cite{spliceai}, we inherit the top-k accuracy defined as follows. Suppose the test set has $k$ true splice sites, and we select $k$ positions that have the highest probability calculated by the model. The fraction of these $k$ positions that belong to true splice sites is reported as the top-k accuracy. Top-k accuracy focuses on the probabilities instead of predictions, so the choice of threshold does not affect it.

\textit{AUC-PR}. The Area under the Precision-Recall Curve (AUC-PR) is another metric that is not affected by threshold selection. As the threshold moves from 1 to 0, we can figure out a series of precision and recall. Plotting precision (y-axis) against recall (x-axis) gives a curve, and the area under it is the AUC-PR.

\begin{table}[t]
	\caption{Performance of H-GSI and SpliceAI-10k on Human dataset with fixed and dynamic thresholds.}\label{tab:res}
	\begin{center}
        \resizebox{1\linewidth}{!}{
            \begin{tabular}{l|ccc|ccc|cc}
			\toprule[2pt]
                & \multicolumn{3}{c|}{Fixed Threshold (0.5)} & \multicolumn{3}{c|}{Dynamic Threshold} & & \\
			Methods & Accuracy & Precision & Recall & Accuracy & Precision & Recall & Top-k & AUC-PR \\ 
                \midrule
                SpliceAI-10k & 0.9297 & 0.9195 & 0.8126 & 0.9401 & 0.8442 & 0.8411 & 0.8417 & 0.8743 \\ 
                \midrule
                H-GSI-TCN & 0.9000 & 0.9284 & 0.7319 & 0.9143 & 0.7710 & 0.7728 &   0.7725 & 0.7982 \\ 
                H-GSI-FCN & 0.9274 & 0.9221 & 0.8061 & 0.9331 & 0.8713 & 0.8220 &   0.8302 & 0.8600 \\ 
                H-GSI-LSTM & 0.9706 & 0.9647 & 0.9223 & 0.9717 & 0.9502 & 0.9245 &   0.9266 & \textbf{0.9387} \\ 
                H-GSI-GRU & \textbf{0.9711} & \textbf{0.9664} & \textbf{0.9231} & \textbf{0.9720} & \textbf{0.9513} & \textbf{0.9256} &   \textbf{0.9276} & 0.9359 \\ 
                H-GSI-Transformer & 0.9401 & 0.9647 & 0.8400 & 0.9516 & 0.8708 & 0.8715 &   0.8714 & 0.9139 \\ 
			\bottomrule[2pt]
		\end{tabular}
        }
	\end{center}
\end{table}

\subsection{Numerical Results}
This subsection reports the performance of H-GSI and contenders using five metrics: random accuracy, precision, recall, top-k accuracy, and AUC-PR. Table \ref{tab:res} lists the performance of the SpliceAI-10K and the proposed H-GSI with fixed and dynamic thresholds described in Subsection \ref{predictor}. It is observed that H-GSI equipped with LSTM, GRU, and Transformer achieves remarkable performance, where all metric values are better than those of SpliceAI-10k. The H-GSI equipped with LSTM wins first place on four metrics except AUC-PR, and the performance of GRU is similar to LSTM. Besides, the performance of H-GSI using the dynamic threshold is better than that using the fixed threshold. In summary, these observations demonstrate the effectiveness of the proposed H-GSI. 

Compared to the results of fixed threshold, the precision and recall of dynamic threshold are more closely related to top-k accuracy. The closer the precision is to recall, the closer the thresholds are calculated on the training and test sets. This observation confirms our conjecture made in Subsection~\ref{predictor}.

Another conclusion comes from an interesting observation that the H-GSI equipped with recurrent-based models surpasses one with Transformer, although the Transformer is nearly 20 times larger than other models. This observation may reveal that recurrent-based models may be more powerful at capturing structural information from sequential data with discrete variables.

Notice that TCN has a similar architecture and parameter size as FCN, but performs worse on most evaluation metrics with a considerable gap. TCN is equipped with causal convolution layers so that the model cannot violate the ordering in which we model the data. However, the identification of splice sites depends on both sides, which does not match the assumption and capacity of TCN.

\section{Conclusions, Discussions, and Prospects}\label{conclusion}
In this paper, we proposed a novel framework for identifying gene splicing, named H-GSI. The proposed H-GSI follows the horizon-wise identification paradigm and comprises four components of the pre-processing procedure transforming string data into tensors, the sliding window technique handling long sequences, the SeqLab model, and the predictor. In contrast to existing studies that process gene information with a truncated fixed-length sequence, H-GSI employs a horizon-wise identification paradigm, contributing to higher accuracy and efficiency. The empirical investigations were conducted on real-world Human splicing variants. The experimental results demonstrated the effectiveness of our proposed method.

We gained a deeper insight into the dataset and surprisingly found that the same sequence has different splice sites, known as alternative splicing. This case has no solution via machine learning since a learning model cannot produce different outputs when fed with the same input. A feasible solution is collaborating on knowledge-based reasoning and data-driven learning \cite{abductive}. For example, one may merge these alternative splice sites, expecting the model to give all candidate predictions at once, and then pick out real splice sites based on domain knowledge or do it manually.

Another topic of future research is the explanation of identification models, which usually includes two types of explanations, known as global and local explanations; the former focuses on what determines the location of a splice site, while the latter investigates why the identification model gives such predictions based on a specific instance interpretation. A feasible way is to dig up various splicing mechanisms by arranging the attention maps and the corresponding nucleotide sequence, reflecting which parts of a sequence attract the attention of the identification model. Taking a deep insight into the identification model may help in figuring out the principle of gene splicing and discovering more splicing patterns.

\section*{Acknowledgements}
This research was supported by the Natural Science Foundation of Jiangsu Province (BK20230782).

\appendix 
\section{Proof of the claim of random accuracy } \label{appendix}
This appendix reviews the claim that the expectation of random accuracy is equivalent to weighted accuracy. The proof is provided as follows.

\begin{proof}
From the definition of random accuracy, we have
\begin{align}
	\mathbb E [Acc]
	&=\mathbb E_{N'}\left[\frac{\sum_{i\in P}\mathbb I_i + \sum_{i\in N'}\mathbb I_i}{m+\gamma n}\right]\\
	&=\frac{1}{m+\gamma n}\left\{\sum_{i\in P}\mathbb I_i + E_{N'}\left[\sum_{i\in N'}\mathbb I_i\right]\right\} \\
	&=\frac{1}{m+\gamma n}\left\{\sum_{i\in P}\mathbb I_i + \sum_{N'}p(N')\sum_{i\in N'}\mathbb I_i\right\} \ ,
\end{align}
where $m=|P|$, $n=|N|$, and $N$ is the subscript set of negative labels (non-splice
sites). 

Provided $\gamma n\in\mathbb N$, one has
\begin{align}
	\sum_{N'}p(N')\sum_{i\in N'}\mathbb I_i
	&=\frac{1}{\binom{n}{\gamma n}}\sum_{N'}\sum_{i\in N'}\mathbb I_i \\
	&=\frac{1}{\binom{n}{\gamma n}}\sum_{N'}\sum_{i\in N}\mathbb I_i\times \mathbb I[i\in N']\\
	&=\frac{1}{\binom{n}{\gamma n}}\sum_{i\in N}\mathbb I_i\sum_{N'}\mathbb I[i\in N'] \\
	&=\frac{1}{\binom{n}{\gamma n}}\sum_{i\in N}\mathbb I_i\times \#[i\in N'] \\
	&=\frac{\binom{n-1}{\gamma n-1}}{\binom{n}{\gamma n}}\sum_{i\in N}\mathbb I_i \\
	&=\gamma\sum_{i\in N}\mathbb I_i \ .
\end{align}
Further, we can conclude
\begin{equation}
\mathbb E [Acc]=\frac{\sum_{i\in P}\mathbb I_i}{m+\gamma n} + \gamma\cdot\frac{\sum_{i\in N}\mathbb I_i}{m+\gamma n} \ .
\end{equation}
Let $w_p$ and $w_n$ denote the weights of positive and negative labels, respectively. When $w_p=1$ and $w_n=\gamma $, the expectation of random accuracy is equivalent to weighted accuracy. This completes the proof.
\end{proof}

\bibliography{JMref}
\bibliographystyle{plain}

\end{document}